\documentclass[aps,twocolumn,
amsmath,amssymb]{revtex4-2}
\usepackage{color}
\usepackage{graphicx}
\usepackage{float}
\usepackage{subfigure}
\usepackage{booktabs}
\usepackage{dcolumn}
\usepackage{bm}
\usepackage{amsmath}
\usepackage{accents}
\usepackage{verbatim}
\usepackage{lineno}
\usepackage{tikz,pgfplots}
\usepackage{appendix}
\usepackage{braket}

\newcommand{\beq}{\begin{eqnarray}}
\newcommand{\eeq}{\end{eqnarray}}

\begin{document}
\title{Revealing Liquid-Gas Transitions with Finite-Size Scaling in Confined Systems}

\author{Chong Zha$^{1,2}$}
\altaffiliation{Contributed equally to this work}
\author{Yanshuang Chen$^{3}$}
\altaffiliation{Contributed equally to this work}
\author{Cheng-Ran Du$^{4}$}
\email{chengran.du@dhu.edu.cn}
\author{Peng Tan$^{3,5}$}
\email{tanpeng@fudan.edu.cn}
\author{Yuliang Jin$^{1,2,6}$}
\email{yuliangjin@mail.itp.ac.cn}

\affiliation{$^1$Institute of Theoretical Physics,
 Chinese Academy of Sciences, Beijing 100190, China}
\affiliation{$^2$School of Physical Sciences, University of Chinese Academy of Sciences, Beijing 100049, China}
\affiliation{$^3$Department of Physics and State Key Laboratory of Surface Physics, Fudan University, Shanghai, 200438, P. R. China}
\affiliation{$^4$College of Physics, Donghua University, Shanghai, 201620, P. R. China}
\affiliation{$^5$Institute for Nanoelectronic Devices and Quantum Computing, Fudan University, Shanghai, 200438, P. R. China}
\affiliation{$^6$Wenzhou Institute, University of Chinese Academy of Sciences, Wenzhou, Zhejiang 325000, China}

\date{\today}

\begin{abstract}
The application of an external field often renders empirical criteria for identifying liquid-gas phase transitions ambiguous.  Here, we demonstrate that the finite-size scaling  of the density profile provides a definitive criterion to distinguish 
liquid-gas coexistence from 
a single fluid phase in field-confined systems. 
Our scaling method collapses the density profiles of different system sizes onto a single master curve for a one-phase system, while causing the profiles to intersect at the interface in a two-phase system. We validate this theoretical proposal through experiments and simulations of two model systems: colloidal suspensions under gravity and/or two-dimensional complex plasmas confined by a central potential. Our method is broadly applicable for detecting liquid-gas phase transitions in laboratory systems where external fields are inherent.
\end{abstract}
\maketitle

\def\thefootnote{*}\footnotetext{These authors contributed equally to this work.}

{\bf Introduction.}
First-order liquid-gas phase transitions (LGPTs) occur at pressures and temperatures below the  critical point. Above the critical point, the transformation between a liquid and a gas is characterized not by  a phase transition, but by supercritical crossovers~\cite{fisher1969decay, xu2005relation,  brazhkin2012two, li2024thermodynamic, jin2025supercritical}. Furthermore, the LGPTs are absent in systems with purely repulsive interactions, such as hard spheres, which exhibit only a single fluid phase.  For many systems, particularly those with unknown inter-particle interactions,  it remains a  non-trivial task to determine whether, and when, an LGPT exists.

In experimental studies, the identification of an LGPT has often relied on direct observational criteria.  For instance, in colloidal suspensions under gravity, phase coexistence is confirmed by observing a meniscus between two amorphous, diffusive phases~\cite{ilett1995phase, poon2015colloidal}. When gravity is eliminated through density-matching,  the two-phase coexistence can be inferred from the formation of dense aggregates (``liquid droplets") within a dilute phase~\cite{nguyen2013controlling, poon1999colloid}, or from the appearance of voids (``gas bubbles") within a dense phase~\cite{ito1994void, dosho1993recent}.

External fields are often unavoidable and sometimes even necessary. For instance, a confining potential is required to prevent charged microparticles from escaping in the ground experiment of a complex  plasma, which is a suspension of such particles in a weakly ionized gas~\cite{konopka2000measurement, melzer2010finite}. 
While it is conventionally believed that particles in a complex plasma interact through a purely repulsive Yukawa potential~\cite{fortov2005complex, morfill2009complex}, precluding LGPTs, recent theoretical and numerical studies suggest that LGPTs may indeed be possible due to mechanisms like effective attraction~\cite{kompaneets2007potential, kompaneets2016interparticle, khrapak2006critical, resendes1998formation, tsytovich1997dust, khrapak2001interaction}, or the cohesive field due to the plasma background~\cite{avinash2007mean}. 
However, LGPTs have never been confirmed yet in experiments due to complex factors  such  as the confining field.
Understanding the fundamental phase diagram of complex plasmas requires a reliable criterion for LGPTs in the situation when a confining field  plays a significant role.

\begin{figure*}[th]
    \centering
    \includegraphics[width=\linewidth]{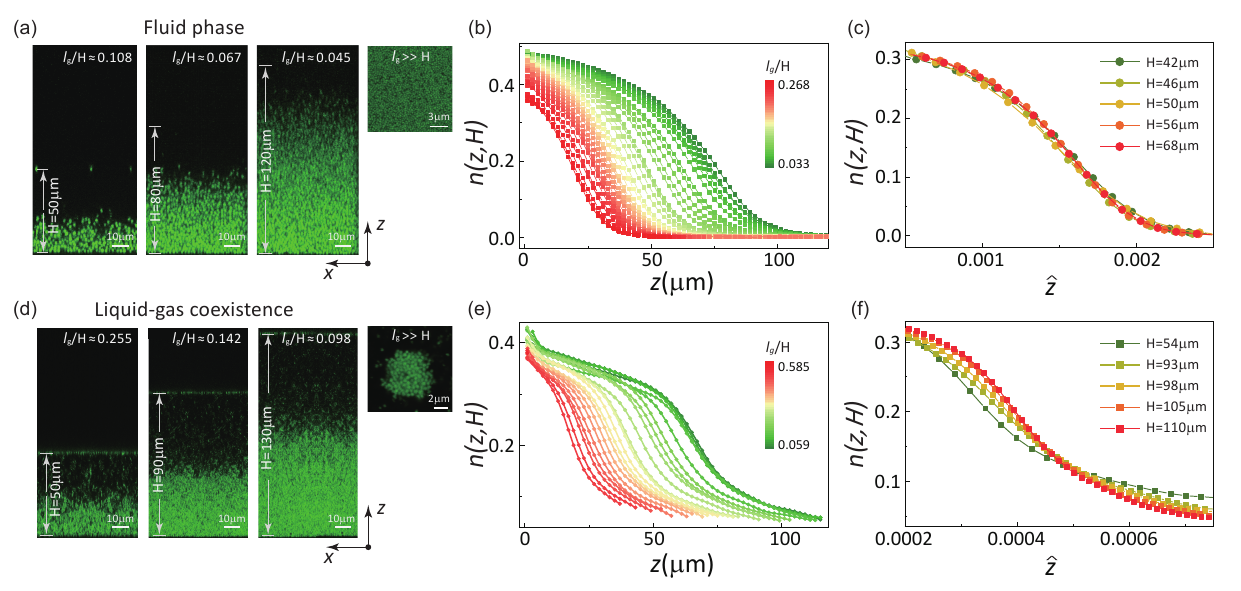}
    \caption{{\bf Results of colloidal experiments.}
    (a-c) Single fluid phase systems with the hard-sphere-like inter-particle interaction. 
    (a) Images of three typical subsystems in the wedge-shaped cell with different $H$ ($\Delta \rho = 0.0472~\rm{g/cm^3}$).
    For comparison, we also show an image of a system under microgravity ($\Delta \rho \approx 0$ and $l_g \gg H$).
    (b) Density profiles $n(z,H)$ of the subsystems with different $H$ in the cell. 
    (c)  $n(z,H)$ as a function of $\hat{z} = z/H$ for different $H$, with a constant {$l_g/H=0.159$}. 
     (d-f) Corresponding results for liquid-gas coexistence systems   where particles interact with additional attraction. 
    }
\label{fig:colloidal_experiments}
\end{figure*}

\begin{figure}[th]
    \centering
    \includegraphics[width=\linewidth]{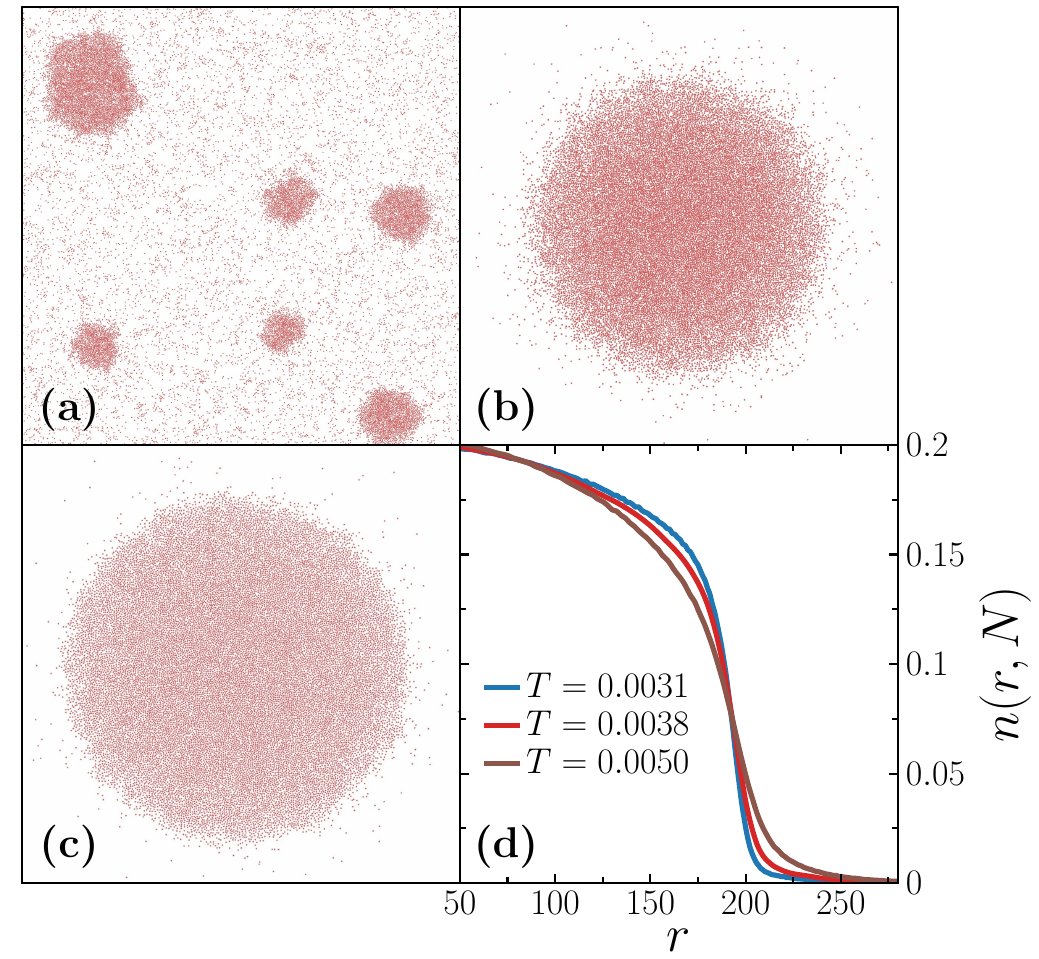}
    \caption{{\bf Simulated complex plasmas.} Snapshots of a Kompaneets system with $T_{\rm c}=0.0038$: (a) without a confining potential, at $T=0.0031<T_{\rm c}$; (b) confined by a central  potential, at $T=0.0050>T_{\rm c}$;  (c)  confined by a central potential, at $T=0.0031$. (d) Density profile of the system in a confining potential at a few different $T$, with $N=20480$. We fix $T/k=6100$.}
    \label{fig:demonstration}
\end{figure}

The presence of an external field can render standard observational criteria for LGPTs misleading. For instance, the droplet/void structures in a zero-field two-phase coexistence system can be removed by a field, by creating a density gradient that suppresses spatial fluctuations; this ambiguity is illustrated  in Fig.~\ref{fig:colloidal_experiments}(a,d) for experimental colloidal systems.  
Another example is given by  a simulated two-dimensional complex plasma  with a liquid-gas critical temperature $T_{\rm c}$ (Fig.~\ref{fig:demonstration}).  
 Below $T_{\rm c}$ and in the absence of a field, the system clearly exhibits phase coexistence, with dense liquid droplets in a dilute gas background (Fig.~\ref{fig:demonstration}a). 
When a central confining field is applied, however, these droplets are driven toward the center, making the visual appearance above and below $T_{\rm c}$ strikingly similar (Fig.~\ref{fig:demonstration}b,c).
More critically, this ambiguity extends to quantitative measures. 
The density profile of a single-phase system can be nearly indistinguishable from that of a system with two-phase coexistence (Fig.~\ref{fig:colloidal_experiments}b,e and Fig.~\ref{fig:demonstration}d).
The density profile evolves smoothly and continuously across $T_{\rm c}$, showing no abrupt change that would signal a phase transition (Fig.~\ref{fig:demonstration}d).

On the theoretical side, the criterion for identifying a phase transition is one of the central questions in statistical physics. A phase transition corresponds to a thermodynamic singularity, which is typically examined through a finite-size analysis~\cite{binder1987finite}. Specifically, near a second-order phase transition at $T_{\rm c}$ (without a field), scale-invariant parameters like the  Binder parameter~\cite{binder1981finite} obey a universal scaling  form,
\beq
B(T,N)= \mathcal{B}(|T-T_{\rm c}|N^{1/d\nu}),
\label{eq:Binder}
\eeq
where  $\nu$ is a  critical exponent, and $d$ the dimensionality. 
This scaling function provides a standard criterion for detecting a second-order phase transition: the phase transition exists if there is  a single intersection point in the curves of $B(T, N)$ versus $T$ for different system sizes $N$. In this study, we generalize such an idea and establish finite-size scaling of the density profile for a system confined by an external field, which can be used to identify a first-order LGPT.

{\bf Theory.}
We consider a system of $N$ particles  confined by a potential $U(r)$ that depends on only one coordinate $r$. 
The ``volume'' is defined by  $V = \int e^{-\beta U(r)}d{\bf r}$, where $\beta$ is the inverse temperature, and  the global (number) density is $\bar{n} = N/V$. 
We discuss the finite-size effects of the density profile $n(r, N)$ in two cases.

{\it (i) Finite-size collapse of the density profile in a single-phase system.} For systems without a phase transition, $n(r, N)$ of different $N$ collapse onto a single curve, as a function of  a rescaled distance parameter $\hat{r} = r/N^{\frac{1}{d}}$:
\beq
n(r,N) = \mathcal{N}(r/N^{\frac{1}{d}}).
\label{eq:scaling_onephase}
\eeq
Equation~(\ref{eq:scaling_onephase}) holds if $\bar{n}$ is kept as a constant for different $N$.
Keeping a constant 
$\bar{n} = 1/\int e^{-\beta U(\hat{r})}d{\bf \hat{r}}$
means that, if the system size is changed as $N\rightarrow\alpha N$,  $U(r)$ should be correspondingly altered by $U(r) \rightarrow U(r/\alpha^{\frac{1}{d}})$.
Note that $d$ is the relevant dimensionality for the change of $N$: e.g., if $N$ is varied in a system under gravity with the cross section $A$ fixed, then $d=1$.

To understand the invariant form Eq.~(\ref{eq:scaling_onephase}), it is instructive to consider the simplest system, an ideal gas, for which $n_{\rm IG}(r, N) = \bar{n}(N) e^{-\beta U(r, N)}$.
This expression becomes $N$-independent if it can be written as, 
$\mathcal{N}_{\rm IG}(\hat{r}) = \bar{n} e^{-\beta U(\hat{r})}$,
where the constant $\bar{n}$ is nothing but a normalization prefactor for the condition $\int n_{\rm IG}(\hat{r}) d{\bf \hat{r}} = 1$.
For non-ideal-gas systems, the simple expression of $\mathcal{N}_{\rm IG}(\hat{r})$ does not hold anymore, due to  the interactions between particles, $u_{\rm int}(r_{ij})$. Nevertheless, the scaling form Eq.~(\ref{eq:scaling_onephase}) is still valid, because (in the thermodynamic limit) the free-energy density $f(\bar{n},T)$ only depends on two thermodynamic parameters, $\bar{n}$ and $T$ (see Supplementary Material for a proof). 
If $\bar{n}$ is $N$-independent, then the free-energy $f(\bar{n},T)$, and consequently the equation of state, are also independent of $N$.

{\it (ii) System-size independence of the density profile at the interface of a two phase coexistence system.} If liquid and gas phases coexist, an interphase is formed between them. The width $W$ of the interphase depends on many factors, such as  thermodynamic parameters ($\bar{n}$ and $T$) and potentials ($U(r)$ and $u_{\rm int}(r)$), but it should not depend on the bulk size $N$. The density profile $n_{\rm interface}(r, N)$ in the interphase region $|r-r_{\rm c}|<W/2$  should collapse for  different $N$: 
\beq
n_{\rm interface}(r, N) = \mathcal{N}_{\rm interface}(r-r_{\rm c}),
\label{eq:scaling_interface}
\eeq
where $r_{\rm c}$ is the center of the interphase. In other words, changing the system size $N$ only shifts the position of the interphase $r_{\rm c}$, but not its width $W$. 

We can follow the idea of Eq.~(\ref{eq:scaling_onephase}) and plot $n(r,N)$ as a function of  the rescaled distance $\hat{r}$. Then Eq.~(\ref{eq:scaling_twophase}) suggests that, 
\beq
n(r, N) = \mathcal{N}[(\hat{r}-\hat{r}_{\rm c}) N^{1/d}].
\label{eq:scaling_twophase}
\eeq
According to Eq.~(\ref{eq:scaling_twophase}), $n(r,N)$ of different $N$ should intersect at a single point: $\hat{r} = \hat{r}_{\rm c}$. This form looks very similar to the universal scaling of the Binder parameter near a critical point, Eq.~(\ref{eq:Binder}).
However, one should keep in mind that the considered liquid-gas transition is a first-order phase transition, and the invariance of  the density profile at $\hat{r}_{\rm c}$ is due to the property of the interface, rather than that of criticality.

In a short summary, we propose to distinguish the liquid-gas phase  transition from a smooth  liquid-gas crossover using the following criterion: plotted as a function of $\hat{r} = r/N^{\frac{1}{d}}$, if  $n(r,N)$ curves of different $N$ intersect at a single point $\hat{r}_{\rm c}$, then the system has a phase transition; if they collapse onto a master curve, then there is no phase transition. 

{\bf Experiments of colloidal suspensions under gravity.} 
The colloidal suspensions are under an effective gravity field ($d=1$) in the the $z$-direction,
$U(z) = m \tilde{g} z$, where $m$ is the particle mass and the effective gravity $\tilde{g} = \gamma g$ is expressed in the unit of the gravity of Earth $g$. According to the above theoretical analysis, the density profile $n(z,N)$ should be plotted as a function of $\hat{z} = z/N$ to examine the finite-size scaling, with  $N \tilde{g}$ fixed in order to keep a constant $\bar{n}=N m \tilde{g}/(k_{\rm B}TA)$.

In the experiment, we study two colloidal systems (see Appendix A). The first system
consists of colloidal particles interacting via  hard-sphere-like, pure repulsive interactions.   This system is expected to exhibit a single fluid phase. Before discussing the effects of gravity, let us first explain the situation under  microgravity condition. The effective gravity $\tilde g \propto \Delta\rho g$ is controlled by the density difference $\Delta\rho$ between 
the particles and the solvent.
If the densities are  matched ($\Delta\rho \approx 0$), then the system exhibits a uniform density field along the $z$-direction with $\tilde g \approx 0$. This is conventionally characterized by  a gravitational length $l_g =\frac{6k_{\rm B} T}{g \Delta \rho (2a)^3}$ that is much larger than the sample height $H$, where $a$ is the particle radius. Such a uniform field for a density-matched system is visualized in the rightmost panel of Fig.~\ref{fig:colloidal_experiments}a. By varying control parameters such as the colloidal volume fraction $\phi$, we do not observe inhomogeneous structures of dense droplets or empty voids. These observations confirm that this system does not have a LGPT, as expected.  

\begin{figure}[th]
    \centering
    \includegraphics[width=\linewidth]{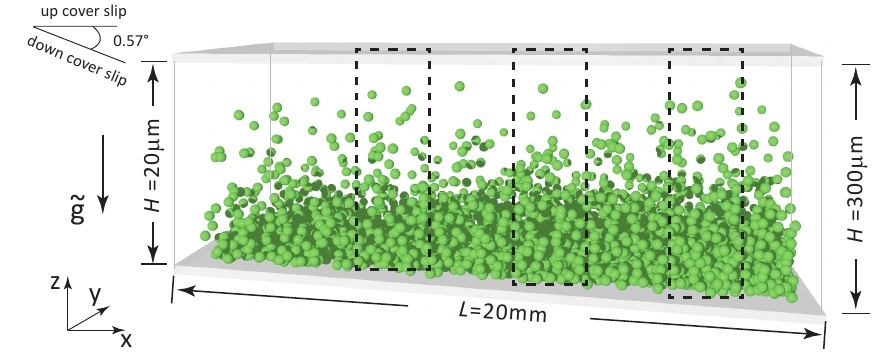}
    \caption{{\bf Schematic of the wedge-shaped sample cell.}   
    We show three typical windows where images are taken (dahsed lines); such images are shown in Fig.~\ref{fig:colloidal_experiments}(a,d).
    }
    \label{fig:exp_setup}
\end{figure}

Next we turn on the effective gravity by mismatching the densities ($\Delta\rho \neq 0$).
To examine the above theoretical scalings, we need to change the system size $N$, and adjust $\tilde{g}$ accordingly to keep $\bar{n} \propto N \tilde{g}$ constant. 
To systematically vary $N$, we prepare wedge-shaped sample cells with a sufficiently small wedge angle of $0.57^\circ$ that
produces a weak, linear dependence of the height $H$ on $x$ (see Fig.~\ref{fig:exp_setup}).
We take $xyz$ 3D image stacks of a subsystem within a volume of 100~$\rm{\mu m\times100~\mu m\times240~\mu m}$. Because the wedge angle is small, the ratio between $H$ and $x$ is only 1.4\%, which makes $H$ vary less than 1.4~$\rm{\mu m}$ inside the subsystem - thus $H$ can be  considered as a constant inside such a subsystem.
Fig.~\ref{fig:exp_setup} shows 2D images of three typical subsystems with different $H$. Importantly, they are aligned along the wedge angle.
Furthermore, the effective gravity $\tilde g \propto \Delta\rho g$ is varied by tuning the density of the solvent.

According to the above setup, the number of particles inside the subsystem is proportional to $H$, $N \propto H$. Thus subsystems with different $N$ (under a given $\tilde{g}$) can be effectively explored by choosing different windows to take the image. 
In addition, the ratio  $H/l_g \propto N \tilde{g} \propto \bar{n}$ controls the global density $\bar{n}$. Therefore we can pick subsystems with proper combinations of $H$ and $\Delta \rho$ such that $\bar{n}$ is kept as a constant -- such subsystems fulfill the requirement of our finite-size scaling analysis proposed in the above theory. In short, $H$ represents the system size, $\Delta\rho$ controls the field, and $l_g/H$ determines $\bar{n}$.

We find that, when $\Delta\rho = 0.0472~\rm{g/cm^{3}}$ ($l_g \approx6.7~{\rm\mu m}$, $0.0335 \leq l_g/H \leq 0.335$), the density field exhibits observable $z$-dependence. 
The first three panels of Fig.~\ref{fig:colloidal_experiments}a visualizes the  density fields for three different $H$ at this $\Delta\rho$, and Fig.~\ref{fig:colloidal_experiments}b shows the corresponding  density profiles  $n(z, H)$ of
this system.
Compared to the microgravity system (the rightmost panel of  Fig.~\ref{fig:colloidal_experiments}a), the density gradient in the $z$-direction is  obvious. 
Interestingly, if we pick $n(z, H)$ curves with different $H$ but the same $l_g/H$, and plot them together as functions of $\hat{z} = z/H$, then these curves fully collapse, following the theoretical scaling Eq.~(\ref{eq:scaling_onephase}) (see Fig.~\ref{fig:colloidal_experiments}c). Thus, the finite-size scaling analysis of experimental density profiles under gravity verifies that hard-sphere-like colloids exhibit a single fluid phase.

For comparison, we prepare a second colloidal system with effective inter-particle attraction induced by the depletion effect~\cite{asakura1954interaction}. 
The phase behavior of this  system is controlled by  $\varphi$, and the polymer concentration $c_p$ that is analogous to the temperature $T$ in the phase diagram~\cite{poon2015colloidal}. Here $c_p$ determines  the strength of the inter-colloidal attraction. 
The existence of a LGPT is identified in the following procedure under the density matched microgravity environment.  At  $\phi=30\%$, we gradually decrease  $c_p$ and  find that at 
$c_p \approx 0.15 c_p^*$, coexistence of liquid-gas states appears, evidenced by the formation of dense (liquid) droplets in a dilute (gas) background (see the rightmost panel of Fig.~\ref{fig:colloidal_experiments}d). Here  $c_p^*$, which is the unit of $c_p$,  represents the overlap concentration at which polymers start to overlap with each other.

Once we turn on the effective gravity by a density mismatch ($\Delta\rho \approx 0.2184~\rm{g/cm^{3}}$), the droplet structure disappears because the particles sediment to form a density gradient in the $z$-direction. 
Typical snapshots of such density fields are shown in the first three panels of Fig.~\ref{fig:colloidal_experiments}d (see also Fig.~\ref{fig:colloidal_experiments}e for the corresponding $n(z,H)$), which are remarkably similar to those of the single-phase system (Fig.~\ref{fig:colloidal_experiments}a,b). Thus gravity makes it highly non-trivial to distinguish between two-phase and one-phase systems, in  sharp contrast to the microgravity situation (the rightmost panels of Fig.~\ref{fig:colloidal_experiments}a,d).
Following the scaling hypothesis Eq.~(\ref{eq:scaling_twophase}), we plot $n(z,H)$ of different $H$ as functions of $\hat{z}$, with the same $l_g/H = 0.085$ (see Fig.~\ref{fig:colloidal_experiments}f). These curves 
 intersect at a point corresponding to a liquid-gas interface, confirming Eq.~(\ref{eq:scaling_twophase}) derived for a two-phase system. 
 To further examine the theoretical scalings, we perform  high-precision molecular dynamics (MD) simulations of colloidal models, and find consistent results (see Appendixes B and C).

\begin{figure}[th]
    \centering
    \includegraphics[width=\linewidth]{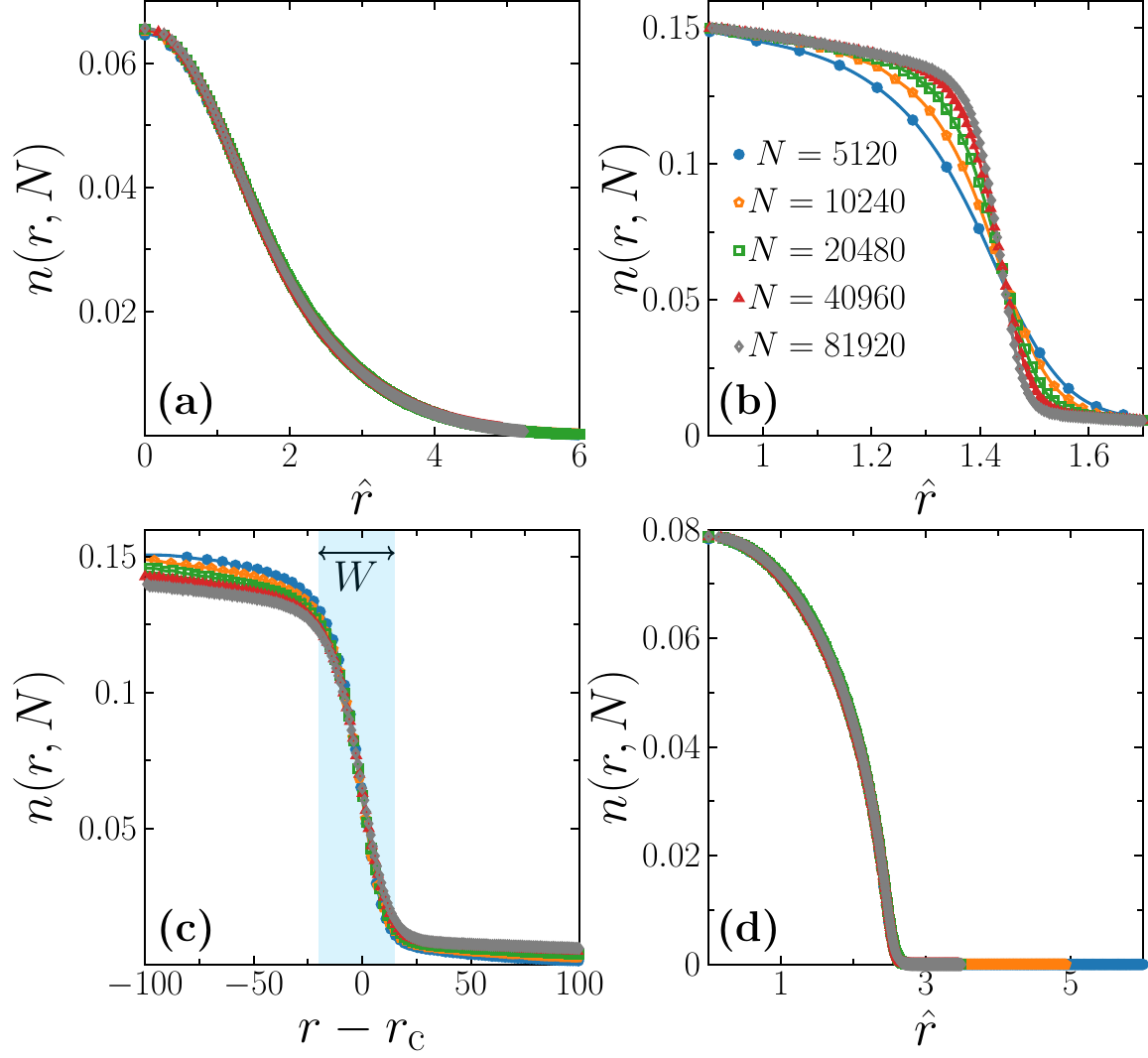}
    \caption{{\bf Density profiles of simulated complex plasmas.}
    (a) A supercritical system with the Kompaneets interaction at $T=0.006$, above $T_{\rm c}=0.0038$. The $n(r, N)$ data of different $N$ are plotted as functions of $\hat{r}=r/N^{1/2}$. (b,c) A subcritical system with the Kompaneets interaction at $T=0.0031$ below $T_{\rm c}$. The intersection point in (b) gives $\hat{r}_{\rm c}=1.44$. In (c), the density profiles are zoomed in around the interface (blue) region. (d) Yukawa systems at $T=0.0031$. We set $Nk=2.62\times 10^{-3}$ for Kompaneets systems and $Nk=2.62\times 10^{-2}$ for Yukawa systems.
    }
    \label{fig:2DComplexPlasmaLorbar}
\end{figure}
{\bf Simulations of complex plasmas in a central confining potential.} Complex plasmas are aerosols composed of a weakly ionized gas and charged microparticles. Because the particles are screened, their interactions are frequently approximated by the Yukawa (Debye-H\"{u}ckel) potential. Since the Yukawa potential is pure-repulsive, the corresponding phase diagram contains only solid and fluid phases - the idealized complex plasmas do not have LGPTs~\cite{fortov2005complex, morfill2009complex}. Under laboratory conditions, the situation becomes much more complicated. First, in order to levitate the particles, a vertical electric field in the sheath region near the lower electrode is introduced to compensate the gravity. In this way, a (quasi) 2D complex plasma is formed in the horizontal plane. The electric field also causes a vertical ion flow, which can result in  effective attraction between particles, as modeled by Kompaneets et al.~\cite{kompaneets2007potential, kompaneets2016interparticle} (Kompaneets potential). Second, the particles are confined horizontally by a central potential, which  pushes the particles towards the center~\cite{konopka2000measurement}. This potential  functions like a ``shallow bowl'' commonly described by a harmonic approximation~\cite{melzer2010finite}. 

In this study, the laboratory  complex plasmas are modeled by a 2D assembly of particles ($d=2$), interacting through the Kompaneets potential, in a central harmonic field $U(r)=\frac{1}{2}kr^2$ (see Appendix D). For the purpose of  finite-size scaling analysis, the specific form of the inter-particle potential and confining potential is not essential. Although other origins of effective inter-particle attraction have been proposed, including other kinds of shadow forces~\cite{khrapak2006critical, resendes1998formation, tsytovich1997dust, khrapak2001interaction}, 
they will not be considered in our model.

In the simulations of the complex plasma system, we choose the parameters $\zeta_l = 0.18$ and $\tau_l = 0$ in the Kompaneets potential. For these parameters, the model has a liquid-gas critical point at $T_{\rm c} = 0.0038$, above which  liquid-gas coexistence does not exist. 
The density profiles $n(r, N)$ are obtained from equilibrium configurations generated by MD simulations.
Typical snapshots of configurations and $n(r, N)$ are shown in Fig.~\ref{fig:demonstration}.
To keep the global density $\bar{n} = \frac{Nk}{2\pi k_B T}$ constant, we fix the value of $N k$.
 The finite-size analysis of $n(r, N)$ is performed for two cases. At $T=0.006$ above $T_{\rm c}$, $n(r, N)$ of different $N$ collapse as a function of $\hat{r}=r/N^{1/2}$ (see Fig.~\ref{fig:2DComplexPlasmaLorbar}a), following Eq.~(\ref{eq:scaling_onephase}). At $T=0.0031$ below $T_{\rm c}$, they intersect at a single point $\hat{r}_{\rm c} = 1.44$ (see Fig.~\ref{fig:2DComplexPlasmaLorbar}b), as expected by Eq.~(\ref{eq:scaling_twophase}). The reason for this invariance is the independence of the interphase on $N$, as described by Eq.~(\ref{eq:scaling_interface}), which is supported by the simulation data (see Fig.~\ref{fig:2DComplexPlasmaLorbar}c). Finally, we validate Eq.~(\ref{eq:scaling_onephase}) by simulation results of (pure-repulsive) Yukawa particles, which should not have a LGPT (see Fig.~\ref{fig:2DComplexPlasmaLorbar}d).

{\bf Conclusion.}
We demonstrate, through colloidal experiments, that our finite-size scaling method can unambiguously identify LGPTs in real systems under confinement. This approach is especially valuable for systems where the confining field is intrinsic. Our simulations pave the way for identifying LGPTs in laboratory complex plasmas. While we focus on two model systems, the method is broadly applicable to  systems with a measurable density profile. Finally, by providing a precise determination of the liquid-gas interface position and width, our method could facilitate advanced studies in interface physics.

\textbf{Acknowledgments.}
Y. Jin acknowledges funding from  Wenzhou Institute (No. WIUCASICTP2022) and the National Natural Science Foundation of China (No. 12447101).
P. Tan acknowledges the National Natural Science Foundation of China (Nos. 12425503, 12174071), the Space Application System of China Manned Space Program (KJZ-YY-NLT0501), the Innovation Program of Shanghai Municipal Education Commission (No. 2023ZKZD06), and Shanghai Pilot Program for Basic Research-FuDan University (No. 22TQ003).
C. Du acknowledges the National Natural Science Foundation of China (No. 12035003).
We acknowledge the use of the High Performance Cluster at Institute of Theoretical Physics, Chinese Academy of Sciences.

\bibliography{LiquidGas}

\clearpage
\centerline{\bf \large End Matter}
\medskip

{\it Appendix A: colloidal experiments - samples, setup and additional data.}
We prepare the following colloidal systems.

(i) Colloidal suspensions with the hard-sphere-like,  pure-repulsive interaction. The suspension consists of PMMA colloidal particles (radius $a\approx 1$~$\mu$m, mass density $\rho\approx 1.2$~$\rm g/cm^{3}$, NBD-dyed and PHSA-grafted, 10\% polydispersity) and a mixed solvent (bromocyclohexane with a density $\rho_1 =1.2~{\rm g/cm^{3}}$, and decahydronaphthalene with a density $\rho_2=0.896~{\rm g/cm^{3}}$). A sufficient amount of salt TBAB (250~$\mu {\rm mol/L}$) is added to the solution to screen the charged interactions between particles.

(ii) Colloidal suspensions with effective attraction due to the depletion effect. 
The suspension consists of colloidal particles ($a \approx 0.3~{\rm \mu m}$, 2.5\% polydispersity), a mixture of bromocyclohexane and decahydronaphthalene, and non-adsorbing polystyrene polymers (molecular weight 2$\times10^6$,  gyration radius  approximately 75~$\rm{nm}$). 
Effective inter-particle attraction is induced by the depletion effect, as explained by the Asakura-Oosawa model~\cite{asakura1954interaction}.  
The dimensionless range  $\xi$ of the effective attractive interaction  is about $0.5 a$. 

The wedge-shaped sample cell is shown schematically in Fig.~\ref{fig:exp_setup}.
The  colloidal suspensions are observed in 3D using a Leica SP8 confocal microscope, with a $z$-direction scanning speed of 10~$\mu$m/s that is much faster than the particles' Brownian diffusion. We use a 40$\times$, 1.3-NA objective lens with a working distance of 240~$\mu$m, which is larger than the height of the sample.

\begin{figure}[th]
    \centering
    \includegraphics[width=\linewidth]{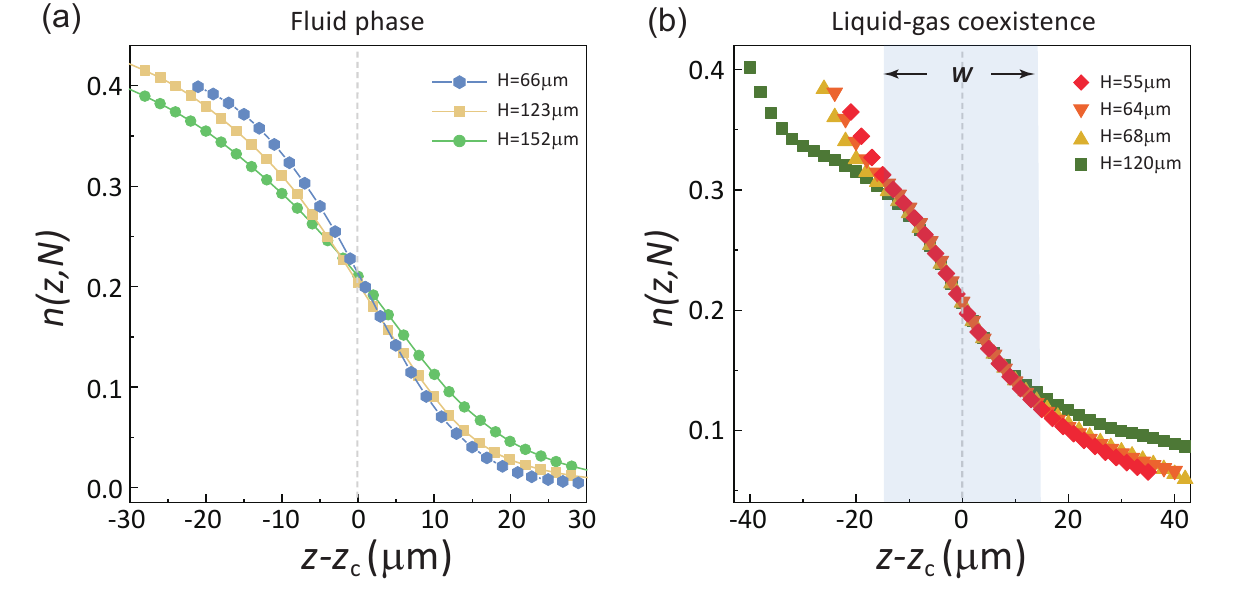}
    \caption{{\bf Density profiles of experimental colloidal suspensions near $z_{\rm c}$.}
    (a) Data for hard-sphere-like colloids, where $z_{\rm c}$ is the position at half height. (b) Data for colloids with effective attraction, where $z_{\rm c}$ is the intersection point determined in Fig.~\ref{fig:colloidal_experiments}f.
    }
\label{fig:ExpIndependenceOfN}
\end{figure}

The density profiles $n(z, H)$ of the two systems are amplified around the center in Fig.~\ref{fig:ExpIndependenceOfN}, for both colloidal systems. As expected, the shifted $n(z, H)$ of different $H$ collapse around $z_{\rm c}$ in the system with effective attraction, suggesting formation of a liquid-gas interface (Fig.~\ref{fig:ExpIndependenceOfN}b). 
\\

{\it Appendix B: simulation model and method of colloidal suspensions.}
For the colloid–polymer mixtures studied in the experiments, the effective interaction between colloidal particles should be regarded as a superposition of hard-sphere-like repulsion and the depletion-induced attraction. In the MD simulations, the steep repulsion part ($0<r<2.0$) is described by a Lennard-Jones (LJ) potential, 
\beq
u_{\rm LJ}(r)=4\epsilon\left[ \left(\frac{\sigma}{r}\right)^{12}-\left(\frac{\sigma}{r} \right)^6 \right],
\label{eq:LJ}
\eeq
with $\sigma=2$ and $\epsilon=1$.
The attractive part is given by the Asakura-Oosawa (AO) model~\cite{asakura1954interaction}
\beq
u_{\rm AO}(r)=- \frac{3\eta_p^{(R)}k_{\rm B}T_{\rm bath}}{2}\left(\frac{1+\xi}{\xi ^3}\right)\left(\frac{r}{2a}-1-\xi \right)^2,
\label{AO_Potential}
\eeq
where $\eta_p^{(R)}$ is the volume fraction of polymer coils, $k_{\rm B}$ the Boltzmann constant, $T_{\rm bath}$ the bath temperature, $a$ the radius of colloidal particles, and $\xi$ the dimensionless range of attraction.
Note that the temperature $T_{\rm bath}$ here does not correspond to the temperature $T$ of colloidal particles, but rather to the ambient (bath) temperature, which is typically taken as the room temperature. We set $\eta_p^{(R)}= 0.06$, $k_{\rm B} = 1$, $a = 1$, $\xi = 0.8$ and $T_{\rm bath} = 1$. 
The particles are placed in a square cylinder,  with elastic boundary conditions at the top and bottom, and periodic boundary conditions on the four side surfaces.
The area of the cross section is $A=120 \times 120$.
The height of the cylinder is sufficiently large such that particles rarely collide with the top boundary, i.e., they are confined by the gravity potential $U(z)=m \gamma g z$, where $m=g=1$. All physical quantities are expressed in LJ units.

We perform MD simulations of isothermal systems using the Large-scale Atomic/Molecular Massively Parallel Simulator (LAMMPS)~\cite{thompson2022lammps}.
The Verlet algorithm with a time step of 0.001  is used to integrate the equations of motion. 
To observe liquid-gas coexistence, the temperature of the colloidal particles is fixed at $T=0.11T_{\rm bath}$, a value between the melting temperature and the critical temperature. In supercritical simulations, the temperature is fixed at a high temperature $T=0.3T_{\rm bath}$. We conduct an additional set of simulations with pure-repulsive interactions, with the LJ potential truncated and shifted to zero around the minimum; 
in this set of simulations, the temperature is fixed at $T=0.11T_{\rm bath}$.
The Nose-Hoover thermostat is employed to update particle velocities, thereby generating a canonical ensemble in the presence of the external potential. 
Upon equilibration, particle configurations are sampled every 50000 steps to compute average values. 
The density distribution is  binned uniformly along the external field direction ($z$-direction) to obtain the averaged  density profile $n(z,N)$.\\

\begin{figure}[th]
    \centering
    \includegraphics[width=\linewidth]{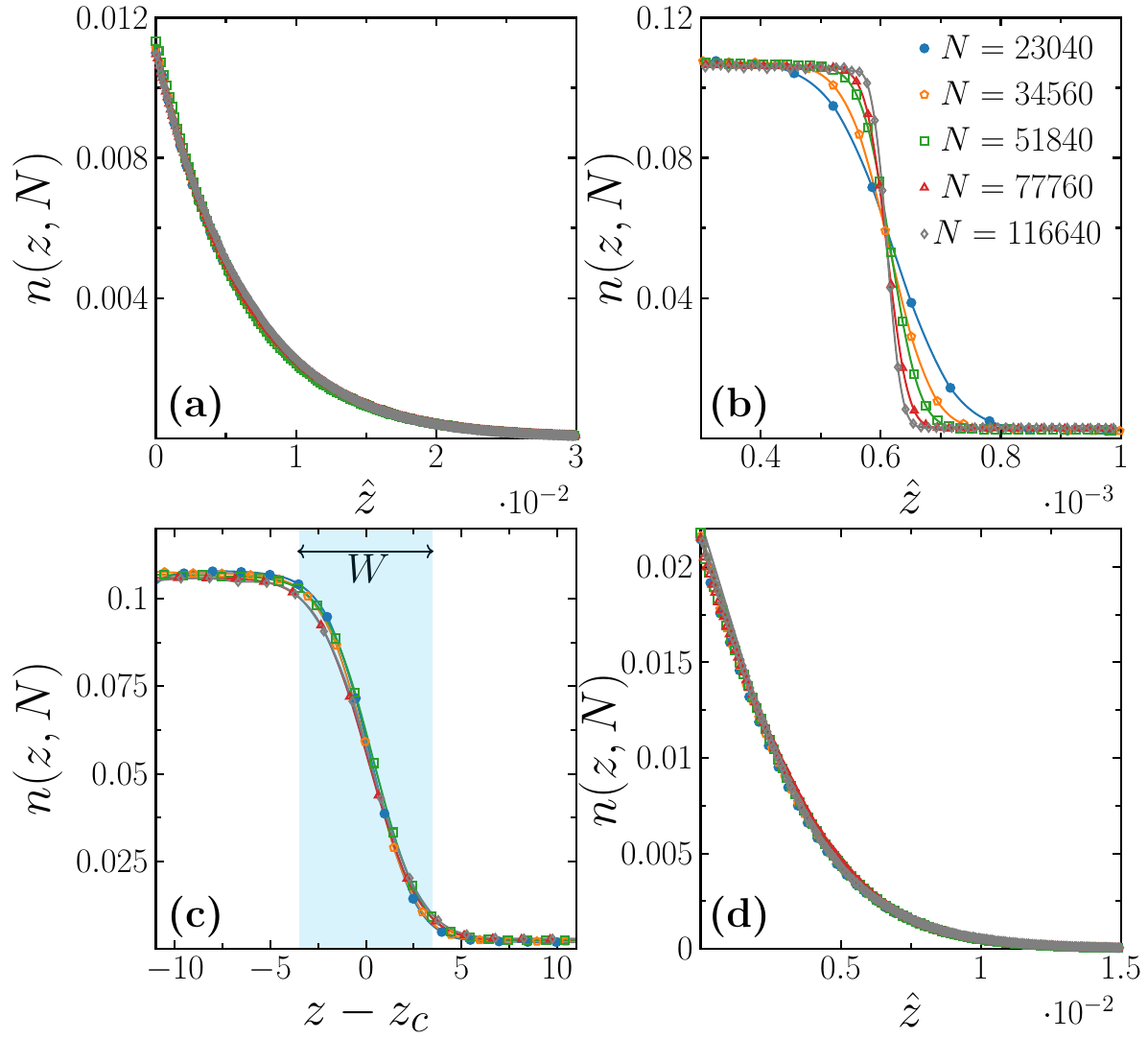}
    \caption{{\bf 
    Finite-size analysis of the density profile $n(z, N)$ in simulated colloidal suspensions in a gravity field.}
    Data in (a-c) are obtained for systems with combined LJ and AO interactions (see Eqs.~\ref{eq:LJ} and~\ref{AO_Potential}), at (a) $T=0.3$ and (b,c) $T=0.11$. Data in (d) are obtained for the truncated LJ potential with the pure-repulsive part, at $T=0.11$.
   Here $\hat{z} = z/N$, and we fix  $N \gamma = 100$.
    }
    \label{fig:3DColloidalSimulation}
\end{figure}

{\it Appendix C: simulation results of colloidal suspensions.} The simulation results are presented in Fig.~\ref{fig:3DColloidalSimulation}. The density profiles $n(z,N)$ are obtained from equilibrium configurations generated by MD simulations. The finite-size analysis is performed for two cases. 
For the case with depletion-induced attraction, at $T=0.3$ above $T_{\rm c}$, the $n(z, N)$ of different $N$ collapse as a function of $\hat{z} = z/N$ (see Fig.~\ref{fig:3DColloidalSimulation}a), following Eq.~(\ref{eq:scaling_onephase}).
To keep $\bar{n} = N \gamma mg/(k_{\rm B}TA)$ constant, we fix  $N \gamma = 100$ for different systems of $N$.
At $T=0.11$ below $T_{\rm c}$, density profile curves intersect at a single point $\hat{z}_{\rm c}=0.00061$ (see Fig.~\ref{fig:3DColloidalSimulation}b), as expected by Eq.~(\ref{eq:scaling_twophase}). The reason for the invariance is the independence of the interphase on $N$, as described by Eq.~(\ref{eq:scaling_interface}), which is supported by the simulation data (see Fig.~\ref{fig:3DColloidalSimulation}c). For the other case with pure-repulsive interaction, the data follow Eq.~(\ref{eq:scaling_onephase}) and show no evidence of a LGPT (see Fig.~\ref{fig:3DColloidalSimulation}d).\\

{\it Appendix D: simulation model and method of  complex plasmas.} The particle interactions are modeled by (i) the Yukawa potential and (ii) the Kompaneets potential \cite{kompaneets2007potential}. The Yukawa interaction is written as,
\beq
\phi_{\rm Y}(r)=\frac{e^{-r}}{r},
\eeq
which is normalized by an energy constant $\frac{Q^2}{4\pi \epsilon_0 \lambda}$, with a Debye length $\lambda =1 $ as the unit of length,  $\epsilon_0$ the permittivity of vacuum, and $Q$ the charge of a single dust particle.

The Kompaneets model involves two independent parameters $\zeta_l$ and $\tau_l$. 
\cite{kompaneets2016interparticle}. The particles are placed on a two-dimensional plane ($z=0$). The interaction is written as,
\beq
\phi_{\rm K}(r)=\frac{2\zeta_l}{\pi}\mathcal{R}e\int_0^{+\infty}dt\frac{K_0\left \{\zeta_l r\sqrt{\frac{t^2+[X(t)+\tau_l]/\zeta_l^2}{1+Y(t)/\zeta_l^2}} \right\} }{1+Y(t)/\zeta_l^2},
\eeq
where
\beq
\begin{cases}
X(t)=1-\sqrt{1+it}, \\
Y(t)=\frac{2\sqrt{1+it}}{it}\int_0^1{\frac{d\alpha}{[1+it(1-\alpha^2)]^2}}-\frac{1}{it(1+it)},
\end{cases}
\eeq
and $K_0(x)$ is the zero-order modified Bessel function of the second kind.  The length unit is the field-induced ion Debye length $\lambda_{E}=\sqrt{\frac{\epsilon_0E_{\rm sh}l}{n_{\rm i}e}}$, 
where 
$E_{\rm sh}$ is the sheath electric field intensity, $n_{\rm i}$ the ion number density, and $e$ the elementary charge. The energy is normalized by $\frac{Q^2}{4\pi \epsilon_0 \lambda_{E}}$.
The interparticle interaction exhibits short-range repulsion and long-range attraction for  certain combinations of $\zeta_l$ and $\tau_l$. In this study, we set $\zeta_l=0.18$ and $\tau_l=0$, for which both short-range repulsion and long-range attraction appear in the potential. 

In the MD simulations, we set up a square simulation box and impose a harmonic confining potential $U(r)=\frac{1}{2}kr^2$, whose center coincides with the center of the square. The box length is chosen to be much larger than the range of the particle distribution, so that the boundary conditions of the box are irrelevant in this simulation. For Yukawa systems, the temperature is chosen as $T=0.0031$. The Kompaneets systems are simulated at a subcritical temperature $T=0.0031<T_c = 0.0038$, and a supercritical temperature $T=0.006>T_c$. The MD simulations are carried out using LAMMPS. The Verlet algorithm with a time step of 0.1  is used to integrate the equation of motion, and the Nose-Hoover thermostat is employed to update particle velocities in a canonical ensemble. After the system reaches equilibrium, particle configurations are sampled every 50000 steps to compute average values. To measure the density profile, we partition the system into concentric annuli of  equal areas,  and compute the number density of particles within each annulus.

\clearpage

\onecolumngrid

\setcounter{figure}{0}
\setcounter{equation}{0}
\setcounter{table}{0}
\setcounter{section}{0}
\renewcommand\thefigure{S\arabic{figure}}
\renewcommand\theequation{S\arabic{equation}}
\renewcommand\thesection{S\arabic{section}}
\renewcommand\thetable{S\arabic{table}}

\centerline{\bf \large Supplementary Material}
\medskip

\section{Partition function of a canonical ensemble in a confining field}

\label{ProofOfLemma1}
We consider a $d$-dimensional system with a fixed particle number $N$, at a temperature $T$, confined by a potential well $U(r)$. Below we use the cluster expansion approach to show that the free energy density of this system, $f=F/N$, depends only on two thermodynamic parameters, $\bar{n} = N/\int e^{-\beta U(r)} d\bf{r}$  and $T$. Note that, due to the confining potential, the pressure $P(r)$ naturally depends on $r$, but we assume that $T$ is $r$-independent. 

The partition function is
\begin{equation}
Z=\frac{1}{N!\lambda^{Nd}}Z_N,
\label{eq:partition function}
\end{equation}
where
\begin{equation}
Z_N=\idotsint d {\bf r}_i \prod_{i=1}^N\left[e^{-\beta U(r_i)}\right] \prod_{i<j}\left[f(r_{ij})+1)\right].
\label{eq:SN}
\end{equation}
Here $f(r_{ij})=e^{-\beta u_{\rm int}(r_{ij})}-1$ is the Mayer function, $u_{\rm int}(r_{ij})$  the interparticle potential, and $\lambda=1$ the de Broglie wavelength.
Following the standard cluster expansion~\cite{mayer1940mayer}, we can write
\begin{equation}
Z_N=\sum_{\{n_l\}}{}^{'}{\prod_{l}{\frac{N!b_l^{n_l}}{n_l!(l!)^{n_l}}}},
\label{eq:clusterExp}
\end{equation}
where $\sum^{'}_{\{n_l\}}$ means a summation over all possible ${n_l}$ under a constraint $\sum_{l=1}^{N}{n_l l}=N$. The cluster integral is,
\begin{equation}
b_l=\idotsint d {\bf r}_i \prod_{i=1}^l\left[e^{-\beta U(r_i)} \right] \sum_{{\rm all\,} l{\rm-clusters}}{ f(r_{ij}) f(r_{mn}) \cdots}
\label{eq:clusterInt}
\end{equation}

As shown in Ref.~\cite{mayer1940mayer}, in the thermodynamic limit, 
\begin{equation}
\lim_{N\rightarrow \infty}{\ln\frac{Z_N}{N!}}=\ln{T_m},
\label{eq:ThermoLimSN}
\end{equation}
where $T_m$ is the largest term in the summation of Eq.~(\ref{eq:clusterExp}).
To calculate $T_m$, we use the method of Lagrange multipliers.  Using the Stirling formula, we can write a general term in  Eq.~(\ref{eq:clusterExp}) as,
\begin{equation}
\ln{T(\{n_l\})}=\sum_{l=1}^N{n_l(\ln{b_l}-\ln{l!}-\ln{n_l}+1)}.
\label{eq:TermInSN}
\end{equation}
The largest term satisfies the conditional extreme value equation:
\begin{equation}
\left. \frac{\partial \ln{T}}{\partial n_l} \right|_{\sum_{l}{n_ll}=N}=0.
\label{eq:ConditionalExt}
\end{equation}
For  $\mathcal{L}=\ln{T}-\eta (\sum_{l}{n_ll}-N)$, the solution of $\frac{\partial \mathcal{L}}{\partial n_l}=0$ is
\begin{equation}
\left\{
\begin{aligned}
n_l &=\frac{b_l}{l!}e^{-\eta l}, \\
N & = \sum_{l=1}^N{\frac{b_l}{(l-1)!}e^{-\eta l}}.
\end{aligned}
\right.
\label{eq:extremumPoint}
\end{equation}
Denoting  $\tilde{b}
_l=\frac{b_l}{b_1}$, we have,
\begin{equation}
\sum_{l=1}^N{\frac{\tilde{b}_l}{(l-1)!}e^{-\eta l}}=\bar{n},
\label{eq:bbarn}
\end{equation}
which means that $x \equiv e^{-\eta}$ can be expressed as a function of $\bar{n}$ and $\{\tilde{b}_l\}$ 
(note that $b_1=\int d {\bf r}e^{-\beta U(r)}$ and  $\bar{n}=\frac{N}{b_1}$).

Finally, in the large-$N$ limit,
\begin{equation}
f = \frac{1}{N}\ln{\frac{Z_N}{N!}}=\frac{1}{N}\ln{T_m}=\frac{1}{\bar{n}}\sum_{l=1}^N{\frac{\tilde{b}_l}{l!}x^l}-\ln{x}.
\label{eq:lnSN}
\end{equation}
We expect  that in this limit, the dimensionless quantity $\tilde{b}_l$ only depends on $\bar{n}$ and $T$, not on $U(r)$. For example, taking $U(r) = \frac{1}{2} k r^2$, we obtain,
\beq
\lim_{N \to \infty} \tilde{b}_2 = \frac{2^{1-d/2}S_d}{\bar{n}} \int_0^\infty r^{d-1} f(r) dr,
\label{eq:b2}
\eeq
where $S_d$ is the surface area of a unit sphere in $d$ dimensions. Equation~(\ref{eq:b2}) does not depend on $U(r)$ because in the $N  \to \infty$, one has to send $k \to 0$, which gives a final expression that depends on the combination of $N$ and $U(r)$, i.e., $\bar{n}$.  Then, according to Eq.~(\ref{eq:lnSN}), 
\beq
f[N \to \infty,T,U(r)] = f(\bar{n}, T).
\label{eq:f}
\eeq
Equation~(\ref{eq:f}) shows that for a system  confined by a field, the proper thermodynamic parameters are $\bar{n}$ and $T$.

\end{document}